\documentclass[10pt]{article} 
\usepackage{amsfonts,amssymb,slashed,makeidx,latexsym,setspace}
\usepackage{graphicx,graphics,floatflt,amssymb,epsf,rotate,subfigure} 
\usepackage{times,cite,color,epsfig,epsf}
\usepackage{multirow}
\usepackage{hyperref}
\def\five{\Phi_5}
\def\fivebar{\Phi_{\bar 5}}
\def\three{\Sigma_3}
\def\eight{\Sigma_8}

\begin{document}

\begin{flushright}
SINP/TNP/2013/14, IPMU-13-0214
\end{flushright}

\begin{center}
{\Large \bf A practical GMSB model for explaining the muon (g-2) \\ with
  gauge coupling unification} \\
\vspace*{1cm} \renewcommand{\thefootnote}{\fnsymbol{footnote}} { {\sf
Gautam Bhattacharyya${}^{1}$},
{\sf Biplob Bhattacherjee${}^{2}$}, {\sf Tsutomu
  T. Yanagida${}^{2}$}, and {\sf Norimi Yokozaki${}^{2}$}
}
\\
\vspace{10pt} {\small ${}^{1)}$ {\em Saha Institute of Nuclear
    Physics, 1/AF Bidhan Nagar, Kolkata 700064, India} \\ ${}^{2)}$
  {\em Kavli IPMU, TODIAS, University of Tokyo, Kashiwa 277-8583,
    Japan} \\ } \normalsize
\end{center}

\begin{abstract}
We present a gauge mediated supersymmetry breaking model having weak
SU(2) triplet, color SU(3) octet and SU(5) 5-plet messengers, that can
simultaneously explain the muon $(g-2)$ data within 1$\sigma$ and the
observed Higgs boson mass of 125 GeV. Gauge coupling unification is
nontrivially maintained. Most of the parameter space satisfying both
is accessible to the 14 TeV LHC.  The lighter of the two staus weighs
around (100-200) GeV, which can be a potential target of the ILC.

\end{abstract}

\setcounter{footnote}{0}
\renewcommand{\thefootnote}{\arabic{footnote}}

\section{Introduction}
Following the latest combination of mass and signal strengths of the
Higgs boson by the ATLAS and CMS collaborations of the Large Hadron
Collider (LHC) \cite{Higgs_ATLAS, Higgs_CMS}, the particle spectrum of
the standard model (SM) is complete and it reigns supreme as an
effective theory for weak scale physics. However, in spite of its
astonishing success, the anomalous magnetic moment of the muon, namely
$a_\mu \equiv (g-2)/2$, remains an enigma.  When compared to the SM
estimate \cite{gm2_hagiwara}, the latest experimental result
\cite{gm2_exp} stands as
\begin{equation}
\Delta a_\mu \equiv (a_\mu)_{\rm exp} - (a_\mu)_{\rm SM} = (26.1 \pm
8.1) \times 10^{-10} \, .
\label{eq:gm2}
\end{equation}
The deviation is above $3\sigma$ level (see also
~\cite{Davier:2010nc}), and it can be resolved if we invoke new
physics at a scale $m_{\rm NP} = \mathcal{O}(100)$ GeV, which follows
from $(\Delta a_\mu)_{\rm NP} \sim (g^2/16\pi^2) (m_{\mu}^2/m_{\rm
  NP}^2) = 20.7 \times 10^{-10} (120\,{\rm GeV}/m_{\rm NP})^2
(g/0.65)^2$, where $g$ is a coupling relevant to the new physics. In
the minimal supersymmetric standard model (MSSM), a resolution of this
deviation requires light superparticles, namely the smuons and
chargino/neutralinos of $\mathcal{O}(100)$~GeV, which propagate in the
loop. With $\tan\beta \equiv \left<H_u\right>/\left<H_d\right> \sim
10$, the size of $\Delta a_\mu$ can be as large as
$\mathcal{O}(10^{-9})$ \cite{gm2susy}. On the other hand, the observed
Higgs boson mass, $m_h \sim 125$ GeV, demands rather large radiative
corrections, which are enhanced by heavy stops weighing
$\mathcal{O}(10)$ TeV or substantial left-right mixing
\cite{OYY}. Gauge mediated supersymmetry breaking (GMSB) models
\cite{GMSB} start with an advantage in this context that at the
supersymmetry breaking scale itself the squarks/guino are heavier than
sleptons/gauginos, i.e. the splitting is in the right direction.
However, in minimal conventional GMSB, which employs a ${\bf 5}$ and a
${\bar {\bf 5}}$ of ${\rm SU(5)}_{\rm GUT}$ as messengers, the heavy
stop pulls up the slepton and weak gaugino soft masses to several
hundred GeV to a TeV which are too high to explain the muon $(g-2)$.

In a previous paper \cite{our_previous} (see also \cite{IMYY}), we
proposed a GMSB model that naturally yielded light uncolored and heavy
colored superpartners. To accomplish this, we employed weak SU(2)
triplet and color SU(3) octet messenger multiplets instead of using
the conventional SU(5) 5-plets. Even with these incomplete SU(5)
multiplets, gauge couplings still unify, though at the string scale
$M_{\rm str} \sim 10^{17}$ GeV which is somewhat higher than the grand
unification theory (GUT) scale $M_G \sim 10^{16}$ GeV. In addition to
satisfying the 125 GeV Higgs boson mass, we could explain the muon
$(g-2)$ at 2$\sigma$ level, with the agreement getting better upon the
addition of SU(5) 5-plet messengers. In the most favorable region the
stau would be the next to lightest supersymmetric particle (NLSP)
being lighter than the bino. For satisfying the cosmological and
accelerators constraints, mild $R$-parity violation (RPV) had to be
invoked which facilitated prompt stau decay.

Recently, radiative corrections to the Higgs mass have been computed
at 3-loop level \cite{H3m} (see also \cite{higgs_3loop}), and it has
been observed that $m_h \sim 125$ GeV is consistent with stop mass as
light as $3-5$ TeV even for minimal left-right scalar mixing
\cite{Feng_3loop}. We show in the present paper that this reduction of
the stop mass allows us to present an improved scenario which is more
comfortable with experimental data.  Through the discussion that
follows, we show that a GMSB model with weak SU(2) triplet, color
SU(3) octet and SU(5) 5-plet messengers not only satisfies $m_h \sim
125$ GeV, but also can explain the muon $(g-2)$ at 1$\sigma$
level. Gauge coupling unification is indeed nontrivially
maintained. No less importantly, we can satisfy the (cosmological)
gravitino problem and the LHC constraints without any need of
introducing RPV operators (For an alternative approach, where
sparticles of 1st/2nd generation are light and of the 3rd generation
are heavy, see \cite{IYY}).

\section{A practical GMSB model}
We employ three types of messenger fields: $\five$($\fivebar$)
transforming as ${\bf 5}$(${\bar{\bf 5}}$) of ${\rm SU(5)}_{\rm
  GUT}$, weak SU(2) triplet $\three ({\bf 1, 3, Y=0})$, and color
SU(3) octet $\eight ({\bf 8, 1, Y=0})$.
The superpotential can be written as
\begin{equation}
W = (M_5 + \lambda_5 F \theta^2) \five \fivebar + (M_8 + \lambda_8 F
\theta^2) {\rm Tr} (\Sigma_8^2 ) + (M_3 + \lambda_3 F \theta^2) {\rm
  Tr} (\Sigma_3^2 ),
\end{equation}
where $F$ characterizes the supersymmetry breaking scale.  The leading
contributions to the gaugino and sfermion masses arising from the
messenger loops are given by
\begin{eqnarray}
m_{\tilde{B}} &\simeq& \frac{\alpha_1}{4\pi} \Lambda_5, \ m_{\tilde{W}} \simeq
\frac{\alpha_2}{4\pi} (2 \Lambda_3 + \Lambda_5), \ m_{\tilde{g}} \simeq
\frac{\alpha_3}{4\pi} (3 \Lambda_8 + \Lambda_5) \, ; \nonumber \\
m_{\tilde{Q}}^2 &\simeq& \frac{1}{8\pi^2} \left[ \frac{4}{3}
  \alpha_3^2 \left(3 \Lambda_8^2 + \Lambda_5^2 \right) + \frac{3}{4}
  \alpha_2^2 ( 2 \Lambda_3^2 + \Lambda_5^2) + \frac{1}{60} \alpha_1^2
  \Lambda_5^2 \right] \, , \nonumber \\ 
m_{\tilde{\bar U}}^2 &\simeq&
\frac{1}{8\pi^2} \left[ \frac{4}{3} \alpha_3^2 (3 \Lambda_8^2 +
  \Lambda_5^2) + \frac{4}{15} \alpha_1^2 \Lambda_5^2 \right] \, ,
\nonumber \\
m_{\tilde{\bar D}}^2 &\simeq& \frac{1}{8\pi^2} \left[ \frac{4}{3}
  \alpha_3^2 (3 \Lambda_8^2 + \Lambda_5^2) + \frac{1}{15} \alpha_1^2
  \Lambda_5^2 \right] \, ,  \\ 
m_{\tilde{L}}^2&=&m_{H_u}^2=m_{H_d}^2
\simeq \frac{1}{8\pi^2} \left[ \frac{3}{4} \alpha_2^2 (2 \Lambda_3^2 +
  \Lambda_5^2) + \frac{3}{20} \alpha_1^2 \Lambda_5^2 \right] \, ,
\nonumber \\
m_{\tilde{\bar E}}^2 &\simeq& \frac{1}{8\pi^2} \left[ \frac{3}{5}
  \alpha_1^2 \Lambda_5^2 \right] \, ; \nonumber
\label{eq:squark}
\end{eqnarray}
where 
\begin{equation}
\alpha_i \equiv \frac{g_i^2}{4\pi}, ~~ \Lambda_8 \equiv
\frac{\lambda_8 F}{M_8}, ~~ \Lambda_3 \equiv \frac{\lambda_3 F}{M_3},
~~ \Lambda_5 \equiv \frac{\lambda_5 F}{M_5} \, .
\end{equation}

\begin{figure}[t]
\begin{center}
\includegraphics[scale=0.9]{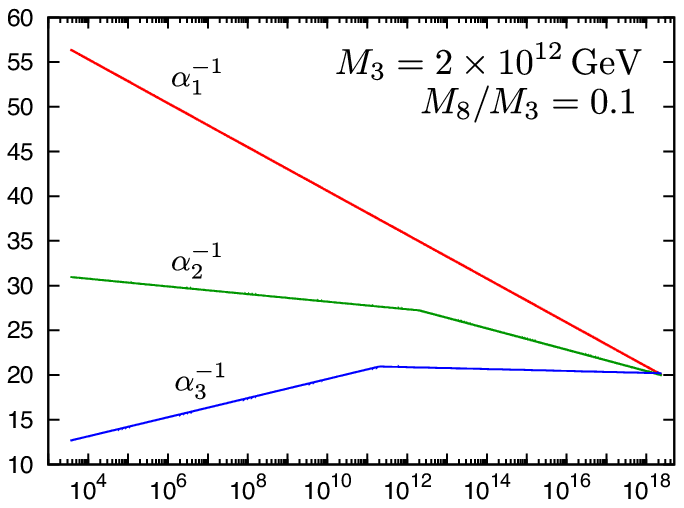} ~~~~
\includegraphics[scale=0.9]{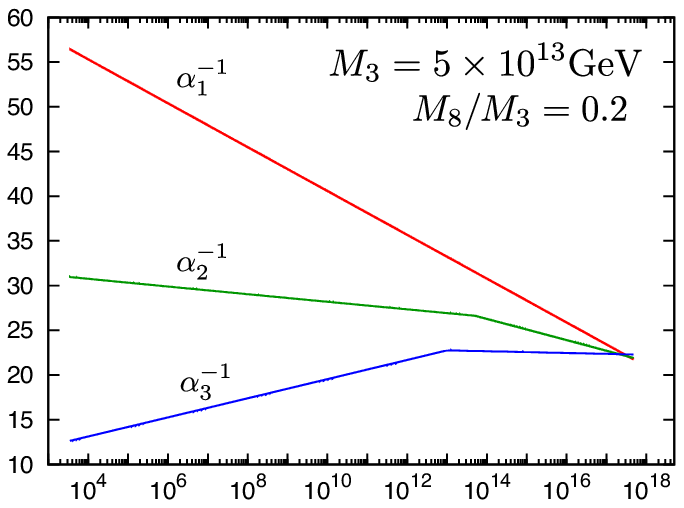}
\caption{\small {\sf Two-loop evolution of the gauge couplings with
    $\three$ and $\eight$ as a function of the renormalization scale
    (GeV).  We take $\alpha_s(M_Z)=0.1184$ and the supersymmetry
    breaking scale $m_{\rm SUSY} \sim m_{\rm stop} \simeq 3.6$ TeV.}}
\label{fig:gut}
\end{center}
\end{figure}

\paragraph{Gauge coupling unification:}
Even with incomplete GUT multiplets, i.e. with $\three$ and $\eight$
only as messengers, gauge couplings do unify with $M_{\rm mess} \equiv
M_8 \sim M_3$ \cite{gut_adjoint}. Solving the coupling evolution
equations explicitly, as done in our previous paper
\cite{our_previous}, it follows that lower the messenger scale $M_{\rm
  mess}$ below the GUT scale $M_G$, higher the actual unification
scale $M_{\rm str}$ above $M_G$. Pushing $M_{\rm str}$ closer to the
Planck scale $M_{\rm Pl} \simeq 2.4 \times 10^{18}$ GeV sets a lower
limit $M_8 \gtrsim 10^{11}$ GeV. The presence of the 5-plets does not
change the evolution slopes, so the above discussion holds in the
present scenario.  In Fig.~\ref{fig:gut}, we exhibit the evolutions of
the gauge couplings with $M_8$ and $M_3$ around $(10^{11}-10^{13})$
GeV scale. The calculation has been performed using the
renormalization group equations (RGE) at 2-loop level
\cite{rge_2loop}. The gauge couplings are unified at $M_{\rm str} \sim
10^{18}$ GeV. This allows us to take $M_8$ closer to its lower
limit. Since $F/M_8$ sets the squark mass, and $F/M_{\rm Pl}$
determines the gravitino mass, a lower value of $M_8$ implies a
lighter gravitino, which helps us solve the cosmological problem (see
later).

\begin{figure}[htbp] 
\begin{center}
\includegraphics[scale=0.9]{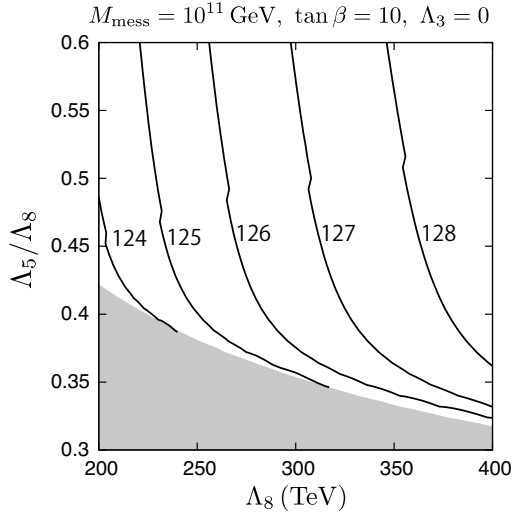}
\caption{\small {\sf Contours of the Higgs boson mass including
    $\mathcal{O}(y_t \alpha_s^2)$ corrections. Here, $m_t=173.2$ GeV
    and $\alpha_s(M_Z)=0.1184$. In the gray region, the stau mass is
    below the LEP2 limit of 90 GeV.}}
\label{fig:mhiggs}
\end{center}
\end{figure}

\paragraph{The Higgs boson mass:}
The observed $m_h \sim 125$ GeV sets the scale of the stop mass, which
in turn fixes $\Lambda_8$.  In Fig.~\ref{fig:mhiggs}, we show the
contours of the Higgs boson mass in the
$\Lambda_8-(\Lambda_5/\Lambda_8)$ plane. The Higgs boson mass has been
evaluated using H3m-v1.2 package~\cite{H3m}, which includes
$\mathcal{O}(y_t \alpha_s^2)$ corrections, where $y_t$ is the top
quark Yukawa coupling.  We take note of the fact that $m_h\sim$ 125
GeV can be explained with $\Lambda_8 \simeq 200-300$ TeV. The
corresponding stop masses are in the $(3.6-5.1)$ TeV range.

\paragraph{Muon $g-2$:} A rough estimate of the Higgsino mixing
parameter is 
\begin{equation}
\mu^2 \sim
(-m_{H_u}^2) \sim \frac{3}{4\pi^2} y_t^2 (m_{\rm stop}^2) \frac{M_{\rm
    mess}}{m_{\rm stop}},
\end{equation}
where $m_{\rm stop}\equiv (m_{\tilde{Q}_3}
m_{\tilde{\bar{U}}_3})^{1/2}$ is the (geometric) average stop mass
scale. For illustration, we have neglected the soft mass of $H_u$
generated at the messenger scale, and considered only the radiative
mass generation. Putting $m_{\rm stop}=3$ TeV and $M_{\rm
  mess}=10^{11}$ GeV, one obtains $\mu \sim 2.7$ TeV.  The value of
$\mu$ is still too large to make the chargino induced contributions to
$(g-2)$ numerically relevant. This contribution is dominated by the
bino-slepton loop, which is given by
\begin{equation}
(\Delta {a_{\mu}})_{\rm SUSY} \simeq
\frac{3}{5}\frac{g_1^2}{8\pi^2}\frac{m_\mu^2 \mu\tan\beta}{M_1^3}
F_b\left(\frac{m_{\tilde{L}}^2}{M_1^2}, \frac{m_{\tilde{\bar
      E}}^2}{M_1^2}\right),
\end{equation}
where $m_{\mu}$ is the muon mass. The contribution is proportional to
the left-right smuon mixing term which contains the $(\mu \tan\beta)$
factor.  The loop function $F_b$ is defined and explicitly displayed
in Ref.~\cite{hagi_gm2_susy} (for a rough guide, $F_b(1,1) = 1/6$).
In order to explain the muon $g-2$, the bino has to be necessarily
light as $\mathcal{O}(100)$ GeV, and the smuon not much heavier.

\begin{figure}[t]
\begin{center}
\includegraphics[scale=0.9]{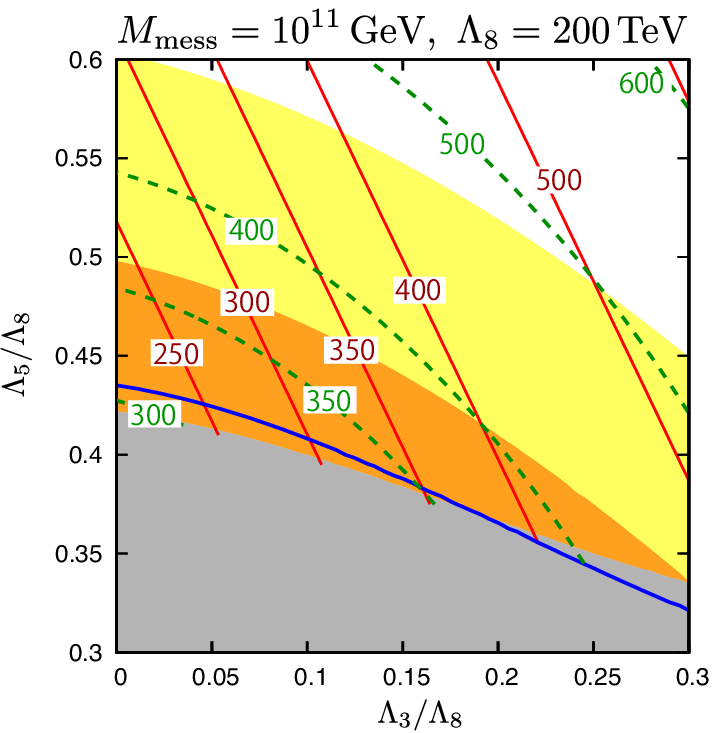} ~~~~
\includegraphics[scale=0.9]{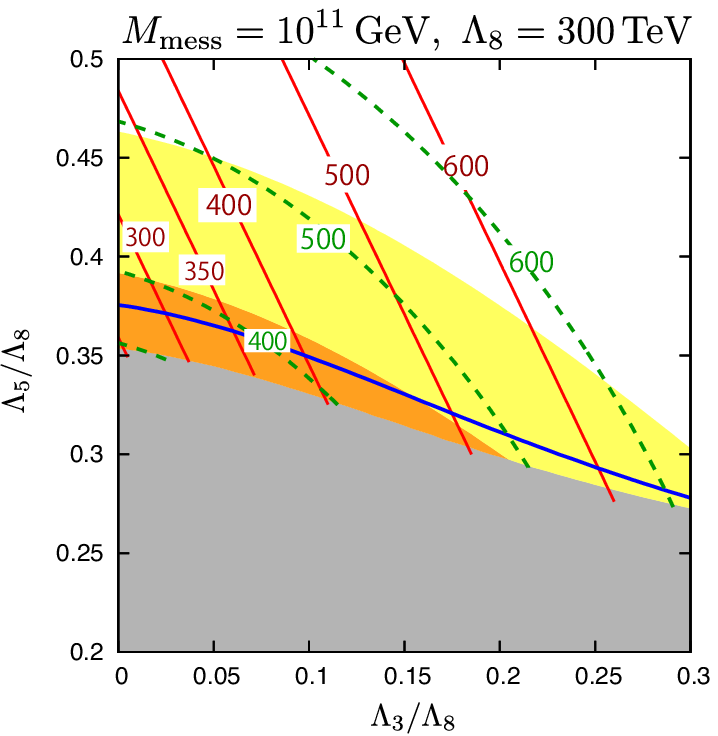}
\caption{\small {\sf In the orange (yellow) region, the muon $g-2$ is
    explained at $1(2)\sigma$ level. The neutralino (stau) is NLSP
    above (below) the blue solid lines. In the gray region, the stau
    mass is smaller than 90~GeV.  The contours of the chargino mass
    (red solid lines) and the soft mass of the left-handed sleptons
    (green dashed lines) are shown in units of GeV.}}
\label{fig:gm2_SUSY}
\end{center}
\end{figure}

In Fig.~\ref{fig:gm2_SUSY} we display to what extent we can explain
the muon $(g-2)$. We take $M_{\rm mess}=M_8=M_3$ for simplicity. The
supersymmetric mass spectrum as well as the RGE running of various
parameters have been performed using {\tt SuSpect} \cite{suspect}. The
supersymmetric contributions to the muon $g-2$ has been evaluated by
{\tt FeynHiggs2.9.5} \cite{FeynHiggs}. To include the threshold
corrections to slepton masses from the Higgsino and heavy Higgs boson
\cite{BPMZ, tata}, we have modified the {\tt SuSpect} package
appropriately.  The contours of different chargino masses have been
shown by red solid lines.  In the orange (yellow) region, the muon
$g-2$ is explained at 1$\sigma$ (2$\sigma$) level, at the same time
keeping consistency with $m_h \sim 125$ GeV.  In the region above the
blue solid line, the neutralino (dominantly the bino, since $\mu$ is
large) is the NLSP, while in the region below the line, the stau is
the NLSP. When the stau is the NLSP, even though it eventually decays
to gravitino, it is stable inside the detector. In this case, the stau
mass of less than 340 GeV is excluded by the LHC data
\cite{stau_stable}. But we need the stau to weigh in the (100-250) GeV
ballpark so that the smuon acquires an appropriate mass to explain the
muon $g-2$ at (1-2)$\sigma$ level. Hence, viable regions are only
above the blue solid line, where the lightest neutralino (dominantly,
the bino) is the NLSP.  Because of the $\Lambda_5$ induced
contributions in Eq.~(3), the bino mass is generated in
a way which is completely uncorrelated to the gravitino mass
generation. The gravitino mass is estimated as
\begin{equation}
m_{3/2} \simeq 0.01\,{\rm GeV} \left(\frac{\Lambda_8}{200 {\rm
    TeV}}\right) \left(\frac{(\Lambda_3/\Lambda_8)}{0.2}\right)
\left(\frac{M_8}{10^{11} {\rm GeV}}\right)
\left(\frac{(M_3/M_8)}{10}\right).
\end{equation}
With this gravitino mass of $\mathcal{O}(10^{-2})$ GeV, the life-time
of the neutralino is $(1-10)$ second, giving a constraint on the
primordial neutralino abundance. However, this abundance is very
small, as a result of which the successful prediction of the big bang
nucleosynthesis (BBN) is maintained \cite{gravitino_problem}, thus
avoiding the gravitino problem.

In Table~\ref{table:mass}, we have presented two sets of reference
points, displaying the mass spectra and the prediction for $(g-2)$,
that pass all constraints. In this context, two types of constraints
deserve special mention: \\
\noindent ($i$)~ When the left-right stau mixing term proportional to
($m_{\tau} \mu \tan\beta$) is large, the charge breaking global
minimum can appear. The life-time of the electroweak vacuum restricts
the size of the $\mu\tan\beta$, which depends of course on the stau
soft mass parameters \cite{stau_vac, stau_Hisano}. However, this
constraint is not very decisive in our case. In fact, the LEP bound on
the stau mass is stronger in the relevant region of the parameter
space \cite{PDG}.

\noindent ($ii$)~ LHC constraints on electroweak gauginos/sleptons
also restrict the relevant parameter space. Searches for three leptons
plus missing energy put a constraint on the wino
mass~\cite{3lepton_LHC}. In the region consistent with the muon $g-2$
at $1\sigma$ level, the left handed sleptons are some what heavier
than the wino. In this case, the final state leptons are the taus
rather electrons/muons, giving the constraint $m_{\chi_1^\pm} \simeq
m_{\chi_2^0} \gtrsim (300-350)$ GeV \cite{3lepton_LHC,
  three_taus}. Note that in some regions of the parameter space, the
wino and the left-handed sleptons are nearly degenerate in mass. These
regions are difficult to be constrained.  Besides, separate (but, not
so tight) constraints exist on the left-handed sleptons, namely,
$m_{\tilde{\ell}_L} \gtrsim 300$ GeV \cite{sleptons_LHC}. The
restrictions on the right-handed sleptons are, however, much less
stringent.

\begin{table}[t!]
  \begin{center}
        \begin{tabular}{  c | c  }
    $\Lambda_3/\Lambda_8$ & 0.17 \\
    $\Lambda_5/\Lambda_8$ & 0.41 \\
    $\Lambda_8$ & 200 TeV \\
    $M_{\rm mess}$ & $10^{11}$ GeV \\
    $\tan\beta$ & 10 \\
    \hline
\hline    
    $\mu$ & $2.4$\,TeV\\
    $m_{\rm stop}$ & 3.6\,TeV \\
    $\delta a_\mu$ & 20.3 $\times10^{-10}$ \\
\hline
    $m_{\rm gluino}$ & 4.4 TeV \\
    $m_{\rm squark}  $ & 4.1 TeV \\
    $m_{\tilde{e}_L} (m_{\tilde{\mu}_L})$ & 379 GeV\\
    $m_{\tilde{e}_R} (m_{\tilde{\mu}_R})$ & 181 GeV \\
    $m_{\tilde{\tau}_1}$ & 123 GeV \\
    $m_{\chi_1^0}$ & 100 GeV \\
     $m_{\chi_1^{\pm}}/m_{\chi_2^0}$ & 375 GeV \\
    \end{tabular}
        \begin{tabular}{  c | c  }
    $\Lambda_3/\Lambda_8$ & 0.11 \\
    $\Lambda_5/\Lambda_8$  & 0.35 \\
    $\Lambda_8$ & 300 TeV \\
    $M_{\rm mess}$ & $10^{11}$ GeV \\
    $\tan\beta$ & 10 \\
    \hline
\hline    
    $\mu$ & $3.5$\,TeV\\
    $m_{\rm stop}$ & 5.1\,TeV \\
    $\delta a_\mu$ & 18.6 $\times10^{-10}$ \\
\hline
    $m_{\rm gluino}$ & 6.3 TeV \\
    $m_{\rm squark}  $ & 5.8 TeV \\
    $m_{\tilde{e}_L} (m_{\tilde{\mu}_L})$ & 425 GeV\\
    $m_{\tilde{e}_R} (m_{\tilde{\mu}_R})$ & 218 GeV \\
    $m_{\tilde{\tau}_1}$ & 133 GeV \\
    $m_{\chi_1^0}$ & 128 GeV \\
     $m_{\chi_1^{\pm}}/m_{\chi_2^0}$ & 411 GeV \\
    \end{tabular}
    \caption{Mass spectra and $(\Delta {a_\mu})_{\rm SUSY}$ for two
      reference points.}
  \label{table:mass}
  \end{center}
\end{table}

\section{Conclusions}
Reconciling the observed Higgs boson mass and the measurement of the
muon $(g-2)$ poses a big challenge to supersymmetric model
building. In this paper we have presented a realistic GMSB model that
can address both these issues satisfying all other constraints.  From
the model-building perspective, the situation has considerably
improved since we constructed the scenario of
Ref.~\cite{our_previous}. We summarize below the salient features
behind this improvement.  The recent 3-loop radiative corrections to
the Higgs boson mass imply that the stop squark is perhaps not as
heavy as order $\mathcal{O} (10)$ TeV. A stop mass of mere (3-5) TeV
can explain the observed 125 GeV mass of the Higgs boson even for
small stop mixing.  A lighter stop means a relatively smaller value of
$\mu$ (but still $\sim$ 3 TeV). This has two implications. First, the
lightest neutralino weighing $\sim$ 100 GeV is bino dominated because
$\mu$ is still quite large ($\sim$ 3 TeV).  Second, the left-right
mixing in the slepton sector, which is proportional to $\mu\tan\beta$,
is relatively smaller because the value of $\mu$ has come down from 6
TeV to 3 TeV thanks to the smaller stop masses.  Since a smaller
left-right mixing implies a smaller splitting between the two slepton
mass eigenvalues of the same flavor, for a wide region of the
interesting parameter space the lightest neutralino can remain lighter
than stau. Note that we could have generated light uncolored and heavy
colored superpartners just with $\three$ and $\eight$, still
nontrivially satisfying gauge coupling unification. The key area where
we really improved with respect to Ref.~\cite{our_previous}, thanks to
the presently known three-loop corrections to the Higgs boson mass and
our assumption of a somewhat {\em late} unification, is that we can
now take bino light enough to explain the muon $(g-2)$ within
1$\sigma$, satisfying at the same time the BBN constraint and LHC data
{\em without} introducing RPV operators. The interesting region of
parameter space can be probed at the 14 TeV LHC, and precision
measurements of the lighter stau can be performed at the future
International Linear Collider (ILC).

\section*{Acknowledgements}
The work of N.Y. is supported in part by JSPS Research Fellowships for
Young Scientists.  This work is also supported by the World Premier
International Research Center Initiative (WPI Initiative), MEXT,
Japan.

\small

\end{document}